\documentclass{article}

\PassOptionsToPackage{square,numbers}{natbib}
\usepackage[final]{nips_2017}

\usepackage[utf8]{inputenc} 
\usepackage[T1]{fontenc}    
\usepackage{hyperref}       
\usepackage{url}            
\usepackage{booktabs}       
\usepackage{amsfonts}       
\usepackage{nicefrac}       
\usepackage{microtype}      
\usepackage{graphicx}
\usepackage{caption}
\usepackage{subcaption}
\usepackage{adjustbox}
\usepackage{amsmath}
\usepackage{xcolor}
\usepackage{siunitx}        
\usepackage[xindy,nomain,acronym]{glossaries}
\usepackage{censor}

\graphicspath{ {images/} }
\StopCensoring
\title{\censor{Now Playing}: Continuous low-power music recognition}

\author{
  Blaise Ag{\"u}era y Arcas\thanks{Authors are listed in alphabetical order by last name.} \,,
  Beat Gfeller,
  Ruiqi Guo,
  Kevin Kilgour,
  Sanjiv Kumar,
  James Lyon,
  \\ \textbf{
  Julian Odell,
  Marvin Ritter,
  Dominik Roblek,
  Matthew Sharifi,
  Mihajlo Velimirovi\'{c}
  }
  \\
  Google Research
  \\
  \texttt{\{blaisea,beatg,guorq,kkilgour,sanjivk,jameslyon,}\\
  \texttt{juliano,marvinritter,droblek,mns,mvelimirovic\}@google.com}
}

\makeglossaries
\newacronym{DSP}{DSP}{digital signal processor}
\newacronym{NNFP}{NNFP}{neural network fingerprinter}
\newacronym{SIFT}{SIFT}{scale invariant feature transform}
\newacronym{FFT}{FFT}{fast Fourier transform}
\newacronym{IIR}{IIR}{infinite impulse response}
\newacronym{DB}{DB}{database}
\newacronym{LSH}{LSH}{locality-sensitive hashing}

\begin{document}

\maketitle
\begin{abstract}
Existing music recognition applications require a connection to a server that performs the actual recognition. In this paper we present a low-power music recognizer that runs entirely on a mobile device and automatically recognizes music without user interaction.
To reduce battery consumption, a small music detector runs continuously on the mobile device's \gls{DSP} chip and wakes up the main application processor only when it is confident that music is present. Once woken, the recognizer on the application processor is provided with a few seconds of audio which is fingerprinted and compared to the stored fingerprints in the on-device fingerprint database of tens of thousands of songs.
Our presented system, \censor{\textit{Now Playing}}, has a daily battery usage of less than \SI{1}{\%} on average, respects user privacy by running entirely on-device and can passively recognize a wide range of music.
\end{abstract}

\section{Introduction}
\label{sec:intro}

When someone is curious about an unknown song that is being played, they currently launch a music recognition application that captures several seconds of audio and uses a server for recognition. Ideally all a curious user would have to do is glance at their mobile device to find out which song is being played without even having to unlock it. There are multiple challenges that have to be tackled in order to make this possible. In this paper we present a first of its kind, continuous low-power music recognizer.

\begin{figure}[t]
\includegraphics[width=\textwidth]{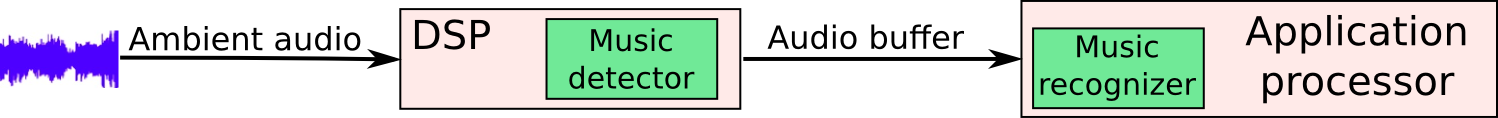}
\caption{\textit{Setup of the continuous music recognition system. The neural network fingerprinter and the fingerprint matcher run on the application processor which is woken up and provided a buffer of audio when the music detector is confident that music is present.}}
\label{fig:overview}
\vspace{-6mm}
\end{figure}

An overview of this setup can be seen in figure~\ref{fig:overview}. First, an audio fingerprinter is required that can run on a mobile device. We developed a neural network fingerprinter to produce compact but discriminative fingerprints from short segments of audio. A small \gls{DB} of fingerprint sequences generated from selected songs is used by a matching algorithm that can quickly match the short sequence of fingerprints from the audio segment.

Because the fingerprinter is relatively computationally expensive and would have a significant impact on the battery if left running all day, a gatekeeper music detector is necessary to avoid triggering it when no music is present. This music detector runs on a separate \gls{DSP} chip and only wakes the main processor when music is present, see Figure~\ref{fig:overview}. Since everything runs locally on the device without sending either audio or fingerprints to a server, the privacy of the user is respected and the whole system can run in airplane mode.

\section{Related Work}
\label{sec:related}

To solve the task of audio content identification, \citet{ouali2016spectrogram} developed a spectrogram-based audio fingerprint which uses multiple binary spectrograms created with various thresholds. Each of the computed fingerprints is then compared against known references. \citet{baluja2007audio} applied computer vision techniques to extract a fingerprint which can be indexed at large scale using \gls{LSH} \cite{gionis1999similarity}.

Another approach that uses binary fingerprints was developed by \citet{tsai2017known}. Their features are derived by applying the constant-Q transform to the audio which is a transformation where the spacing and width of its filters match the pitches of the musical scale.
The fingerprints are then binarized using a hamming embedding.

\citet{malekesmaeili2014local} designed an audio fingerprint which is invariant to time or frequency scaling. Their fingerprint is based on the time-chroma representation of the audio signal which groups all pitches that differ by an octave together. Unlike the logarithmic Mel scale, pitch shifting a song by a full octave will result in the same time-chroma representation and pitch shifts by fractions of an octave only shift the values along the chroma axis. They show that their fingerprints can outperform \gls{SIFT} based fingerprints.

An audio fingerprinting technique explicitly designed for embedded applications is proposed by \citet{plapous2017low} where they forgo the \gls{FFT} that most other fingerprint techniques use in favor of using optimized \gls{IIR} filter banks. They were able to improve both the computational speed of the audio fingerprints as well as reduce their size while maintaining the accuracy of their baseline Robust Audio Hashing technique \cite{haitsma2001robust}.

\citet{chandrasekhar2011survey} present an overview of music identification techniques and compare their usefulness for mobile devices based on a set of noisy audio clips.

A number of popular music identification applications are available for mobile devices such as Shazam \cite{wang2003industrial,wang2006shazam}, SoundHound \cite{soundhound} and Google Sound Search \cite{soundsearch}. These applications are typically manually triggered by users and require a connection to a server for recognition.

\section{Neural Network Fingerprinter}
\label{sec:fingerprinter}

\begin{figure}[t]
\centering
\begin{subfigure}[b]{.39\textwidth}
\vspace{-2mm}
\includegraphics[width=0.7\textwidth]{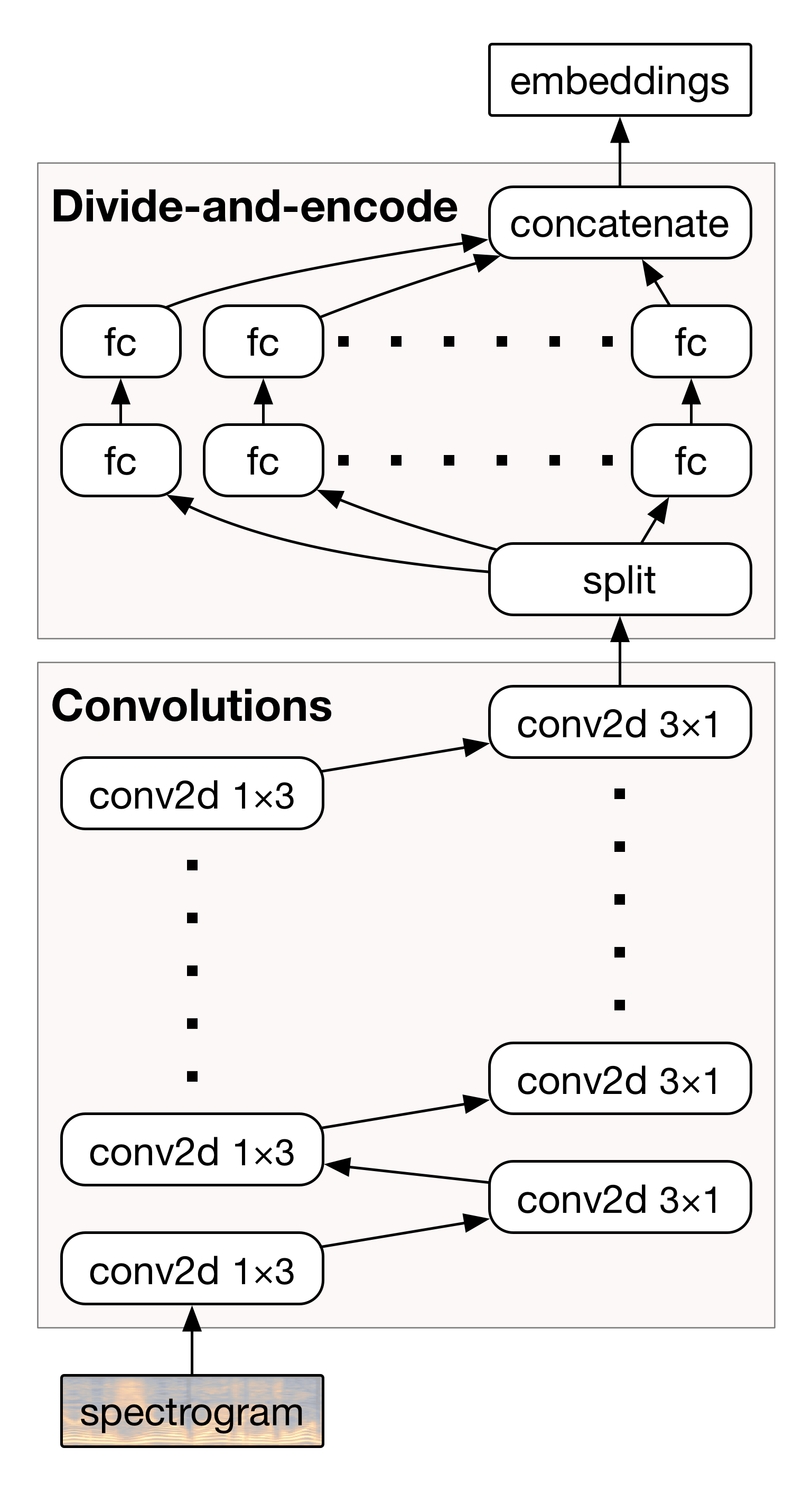}
\caption{\textit{Topology of the music recognizer.}}
\label{fig:nnfp_topo}
\end{subfigure}
\vspace{1ex}
\begin{subfigure}[b]{.5\linewidth}
\includegraphics[width=0.7\textwidth]{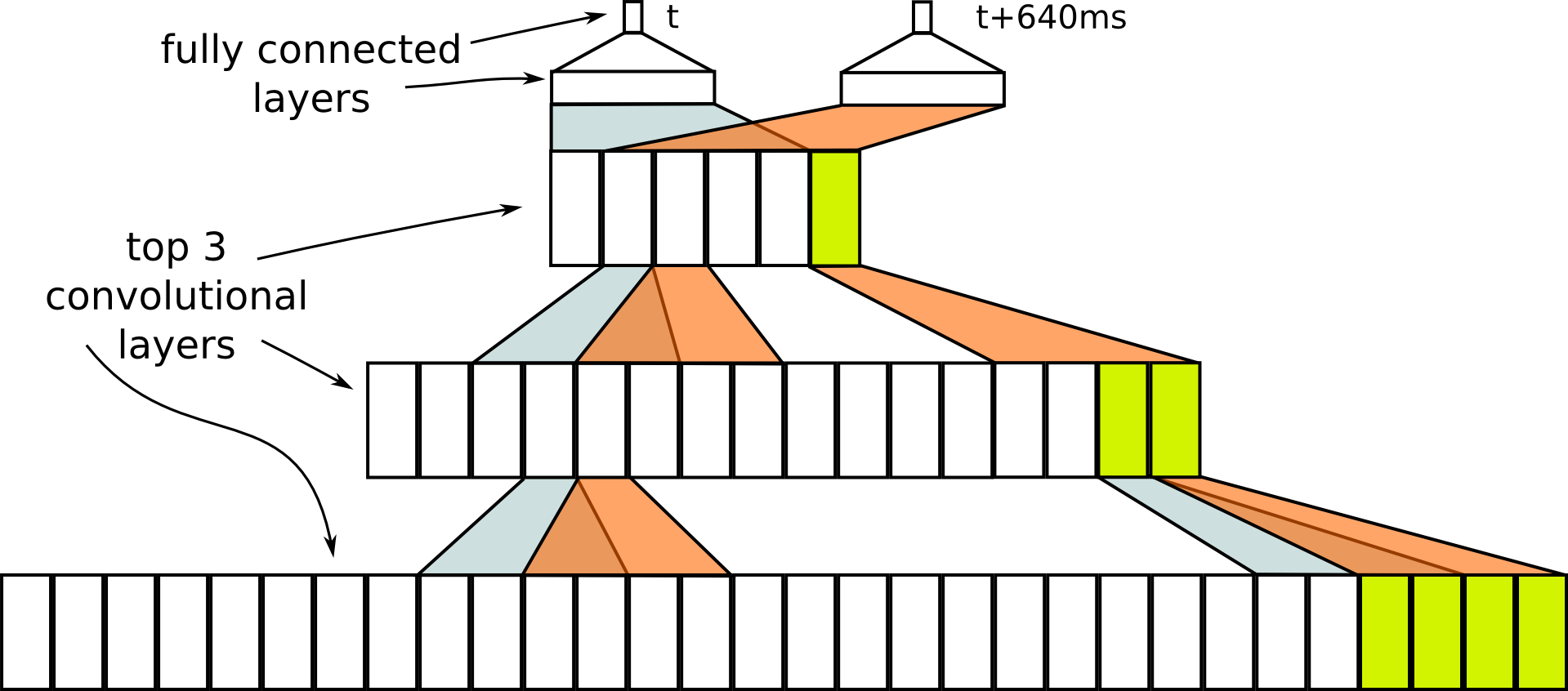}
\caption{\textit{Music detector inference. Highlighted layer vectors are required for the next output at t + \SI{640}{ms}.}}
\label{fig:musdet_inference}
\par\vfill
\vspace{2mm}
\begin{adjustbox}{max width=0.7\textwidth}
\begin{tabular}[b]{c|c|c}
layer & size-in &  kernel \\ \hline
separable conv2d &   446x1x32   &   4x1, 2  \\
separable conv2d &   222x1x32   &   4x1, 2  \\
separable conv2d &   110x1x32   &   4x1, 2  \\
separable conv2d &   54x1x32    &   4x1, 2  \\
separable conv2d &   26x1x32    &   4x1, 2  \\
separable conv2d &   12x1x32    &   4x1, 2  \\
flatten          &   5x1x32     &           \\
fully connected  &   160        &           \\
fully connected  &   8          &           \\
\end{tabular}
\end{adjustbox}
\caption{\textit{Music detector topology. All layers but the last use the ReLU activation function and batch normalization.}}

\label{tab:musdet_topo}

\end{subfigure}
\vspace{-2mm}
\caption{\textit{Architectures of the music detector and recognizer neural networks.
}}
\vspace{-5mm}

\end{figure}

At the heart of our music recognizer is the \gls{NNFP} which analyzes a few seconds of audio and emits a single fingerprint embedding at a rate of one per second. A detailed structure of the \gls{NNFP} can be seen in Figure~\ref{fig:nnfp_topo}. A stack of convolutional layers is followed by a two-level divide-and-encode block \cite{Lai_2015_CVPR} which splits the intermediate representation into multiple branches. All layers except for the final divide-and-encode layer use the ELU \cite{clevert2015fast} activation function and batch normalization.

We trained the network using the triplet loss function \cite{schroff2015facenet}, which for every audio segment minimizes the distance to examples of the same audio segment, while ensuring that their distances to all other audio segments are larger. In this context, audio segments are only considered the same if their starting positions differ by less than a few hundred milliseconds and are from the same song.

The \gls{NNFP} model is trained on a dataset of noisy music segments which are aligned to the corresponding segment in their reference song.

\section{Fingerprint Matching}
\label{sec:matching}

During matching, we need to identify the fingerprints sequence in our song \gls{DB} which most closely matches the sequence of fingerprints generated by the query audio. We perform sequence search in two stages. First, each fingerprint of the query sequence is searched using an approximate nearest neighbor method to find the top-$K$ nearest fingerprints in the \gls{DB}. Then, we perform fine-grained scoring with the most promising candidate sequences.

Due to constraints on energy consumption and storage, we had to compress the \gls{DB} while supporting fast on-device retrieval.
We adopted a strategy similar to \citet{guo2016mips,wu2017nips}, which uses a hybrid solution of trees and quantization where parameters are trained in a data-dependent fashion.

At indexing time, we minimize the quantization error $\lVert x - \tilde{x}\rVert_2$, where $x$ is a fingerprint vector and $\tilde{x}$ is its compressed approximation, as a proxy for minimizing the distance approximation error when searching for query fingerprint $q$: $\left| \lVert q - x\rVert_2 - \lVert q - \tilde{x}\rVert_2 \right|$.

At search time, vector quantizers are used to identify a set of partitions
close to the query. An exhaustive search of those partitions only touches
approximately \SI{2}{\%} of the \gls{DB}. The stored quantized encoding of
the fingerprints is approximately 32 times smaller than the original floating point
representation.

Most errors from searching the subset of compressed fingerprints can be
recovered during the sequence matching stage. This stage reuses promising
partial matches found through an approximate search, and retrieves full
sequences of fingerprints from the \gls{DB}. Our sequence similarity metric
accounts for search errors and uses a method inspired by \citet{qin2013adaptive},
where the similarity between two fingerprints is determined by the local density
of fingerprints in our song \gls{DB} and adjusts the final acceptance threshold
based upon the set of similar sequences found. The effectiveness of this
approach is shown in Figure~\ref{fig:adaptive-scoring}.

\section{Music Detector}
\label{sec:musdet}
The music detector runs continuously on a separate \gls{DSP} chip and acts as a gatekeeper so the computationally expensive fingerprints are only computed when ambient music is playing. Memory and operations per second are extremely limited on the \gls{DSP} chip so as to avoid draining the device's battery. The music detector runs in 3 stages. In the first stage, log Mel features are extracted from the audio stream. 

In the next stage, a neural network computes the probability of music playing using a window of the computed log Mel feature vectors. The network, described in detail in table~\ref{tab:musdet_topo} consists of a stack of convolutional layers, each of which reduces the dimensionality of its input by a factor of two. In total the model has \SI{8}{k} parameters and occupies less than \SI{10}{KB} of memory. The kernel stride of 2 means that each convolutional layer requires 2 new input vectors in order to produce a new output vector. This can be seen in figure~\ref{fig:musdet_inference} which depicts the network's top 3 convolutional layers. The network containing a stack of 6 convolutional layers therefore outputs a new prediction every $2^6$ frames or \SI{640}{ms}.

The music detector neural network is trained on a dataset containing a subset of AudioSet \cite{audioset2017} and an additional set of noise-added audio clips which were labeled with music present or not present. During training we take a random sub-clip as the input for the network.
We simulate quantized activations during training.

In the final stage the stream of predictions from the neural network is averaged over a sliding window of a few seconds. After $c$ consecutive music predictions with a confidence over a threshold $t$, a detection is registered and an audio buffer is sent to the fingerprinter on the application processor. This smoothing and aggregation helps to filter out some of the neural network's errors and ensures that the audio buffer contains a sufficient amount of music to be recognized.

\section{Evaluation}
\label{sec:eval}

\begin{figure}[t]
\begin{subfigure}[t]{0.45\linewidth}
  \centering
\includegraphics[width=0.8\textwidth]{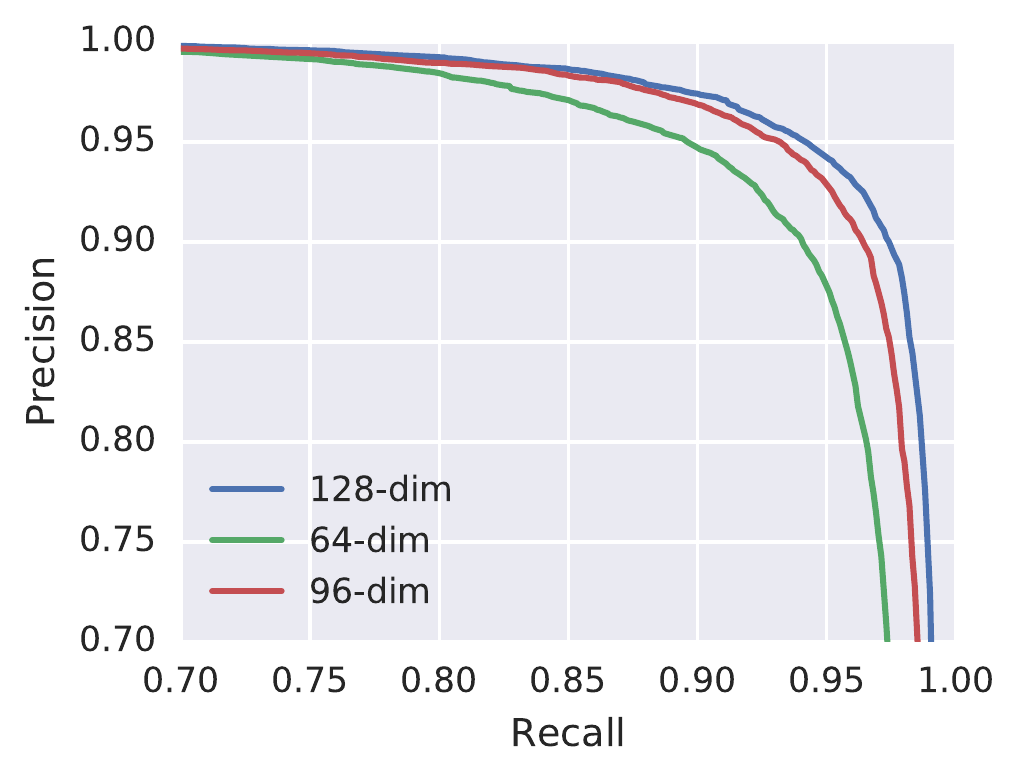}
\caption{Performance of embeddings with different numbers of dimensions. The
models generating these embeddings are identical except for the final divide-and-encode
block.}
\label{fig:embbing-size}
\end{subfigure}
\begin{subfigure}[t]{0.45\linewidth}
  \centering
\includegraphics[width=0.8\textwidth]{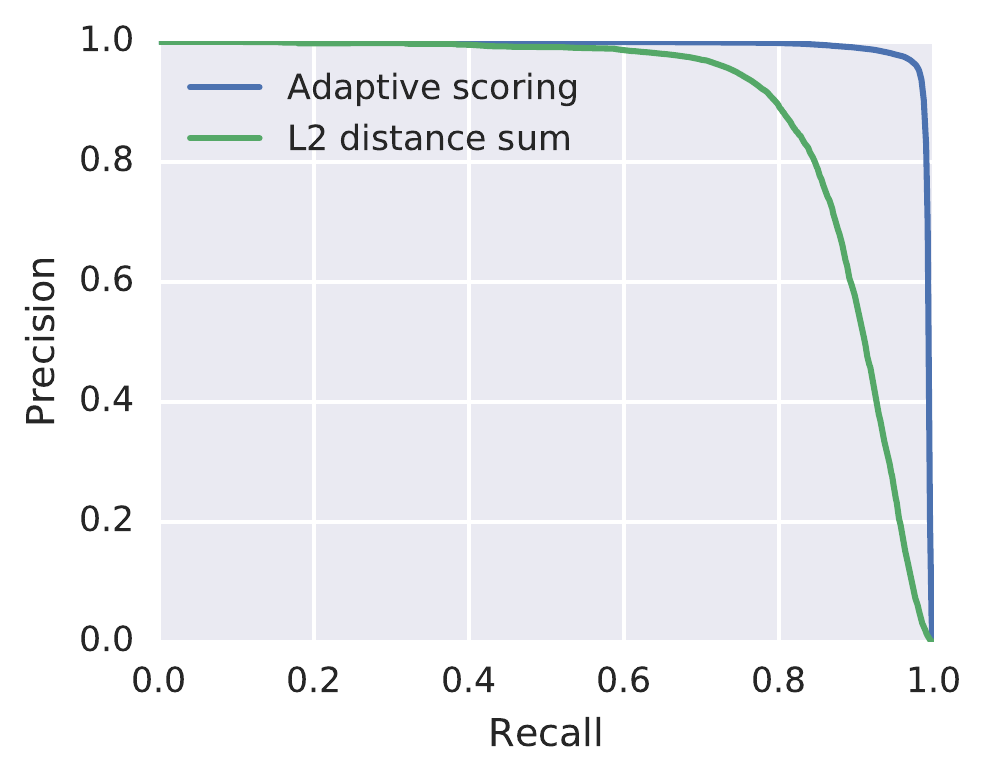}
\caption{Comparison of adaptive scoring similarity metric against a na\"ive
implementation using $L^2$ distance.}
\label{fig:adaptive-scoring}
\end{subfigure}
\vspace{-2mm}
\caption{\textit{Evaluation of the neural network fingerprinter and fingerprint matcher.}}
\vspace{-4mm}
\end{figure}

Given the limited space available on a mobile device for the fingerprint \gls{DB}, we had to find a compromise fingerprint size. Smaller fingerprints allow for more songs in the \gls{DB} while larger fingerprints result in a higher recognition accuracy. We evaluated the performance of the \gls{NNFP} and the matching algorithm using 64, 96 and 128 dimensional fingerprints on a set of \SI{20}{k} \SI{8}{s} long audio segments from \SI{10}{k} different songs. Figure~\ref{fig:embbing-size} shows the precision/recall curves of these three models. The 96 and 128~dimensional models clearly outperform the 64~dimensional model with the 128~dimensional model being only slightly better than the 96~dimensional model. We decided to use the 96~dimensional model so that each song occupies on average less than \SI{3}{KB} in our \gls{DB}, roughly \SI{30}{\%} less than the 128~dimensional model.

\subsection{Music detector performance}
In order to evaluate the performance of the music detector, we tested it on \SI{12}{k} short (\SI{16}{s} - \SI{40}{s}) music regions within a \SI{450}{h} audio test set. The test set contains various background noises such as speech, office noise, street sounds, walking, and vehicle (train, bike, car) noises. The music varied in loudness from imperceivable by humans to very loud. For the music detector we had to make a trade off between wanting a high recall (always triggering with music regions) and avoiding false positives. By accepting a false positive rate of roughly once every 20 minutes on non-silent audio, we were able to still maintain a recall of \SI{75.5}{\%}.

\subsection{Energy consumption evaluation}
On both the \censor{Pixel 2} and \censor{Pixel 2 XL} devices, activating the \gls{DSP} music detector resulted in a small increase of stand-by power consumption of under \SI{0.4}{mA}. Further power is only consumed when the application processor is woken up to identify a song, with each wakeup costing around \SI{2.0}{J} or \SI{0.139}{mAh} which on average occurs about 100 times a day. In total this results in \censor{\textit{Now Playing}} consuming on average \SI{23.5}{mAh} or about \SI{0.9}{\%} of the \censor{Pixel 2}'s \SI{2700}{mAh} battery per day.

\section{Conclusion}
\label{sec:conclusion}
We presented a continuous music recognition system that uses a low-power music detector to only wake up the main application processor only when music is present, thereby keeping the total daily power consumption under \SI{1}{\%} of the battery for an average user.
Using our neural network fingerprinter we are able to generate compact and discriminative fingerprints at a rate of one 96~dimensional embedding per second, which means that each song occupies on average less than \SI{3}{KB} in our on-device \gls{DB}, alleviating the need to query a server and thereby preserving the user's privacy.
The overall system we outlined allows us to detect, recognize and inform users which song is playing without them having to touch their \censor{Pixel 2} phone and without any data leaving the device.

\subsubsection*{Acknowledgments}
The authors would like to thank Gabriel Taubman, Katsiaryna Naliuka, Chris Thornton, Jan Althaus, Brandon Barbello and Tom Hume, all with Google, for their help with the implementation of \textit{Now Playing}. The authors would also like to thank David Petrou, Felix Yu and Corinna Cortes for their thoughtful comments on the paper.

\bibliographystyle{abbrvnat}
\bibliography{references}

\end{document}